\documentclass{IEEEtran}
\usepackage{cite}
\usepackage[utf8]{inputenc}
\usepackage{amsmath}
\usepackage{graphicx}
\begin{document}

\title{Speed-of-sound imaging by differential phase contrast with angular compounding}
\author{Nikunj~Khetan, Timothy~Weber, Jerome Mertz%
	\thanks{This work was supported by National Institute of Health grant R21-GM134216. We thank Rouhui Yang for initial work on this project.}
	\thanks{N. Khetan is with the Mechanical Engineering Department, Boston University, 8 St. Mary's St., Boston MA 02215}
	\thanks{T. Weber and J. Mertz are with the Biomedical Engineering Department, Boston University, 44 Cummington Mall, Boston MA 02215. (e-mail: jmertz@bu.edu)}}

\maketitle

\begin{abstract}
We describe a technique to reveal speed-of-sound (SoS) variations within an echogenic sample. The technique uses the same receive data as standard pulse-echo imaging based on plane-wave compounding, and can be operated in parallel. Point-like scatterers randomly distributed throughout the sample serve as local probes of the downstream transmit-beam phase shifts caused by  aberrating structures within the sample. Phase shifts are monitored in a differential manner, providing signatures of transverse gradients of the local sample SoS. The contrast of the signatures is augmented by a method of angular compounding, which provides ``focus" control of the image sharpness, which, in turn, enables a visual localization of aberrating inclusions within the sample on the fly. The localization can be performed in 2D when operated with standard B-mode imaging, or in 3D when operated with C-mode imaging. Finally, we present a wave-acoustic forward model that provides insight into the principle of differential phase contrast (DPC) imaging, and roughly recapitulates experimental results obtained with an elastography phantom. In particular, we demonstrate that our technique easily reveals relative SoS variations as small as 0.5\% in real time. Such imaging may ultimately be useful for clinical diagnosis of pathologies in soft tissue.  
\end{abstract}            

\section{Introduction}

Pulse-echo sonography is by far the most established modality for clinical ultrasound imaging. Such imaging is generally based on the assumption that the speed of sound (SoS) is uniform throughout the sample. However, it is well known that different tissue types can produce variations in the SoS \cite{bamber1979}. For one, such variations cause aberrations that can degrade pulse-echo image quality \cite{anderson2000}. But more importantly they can themselves be of diagnostic value in differentiating diseased from healthy tissue. For example, SoS imaging by ultrasound computed tomography (UCT) has been successfully applied to disease diagnosis of both liver \cite{lin1987} and breast \cite{jeong2008,li2009}.  However, UCT requires one or more receivers \cite{glozman2010,duric2013,sak2015,zografos2015} or a passive reflector \cite{nebeker2012} to be placed on the distal side of the tissue, limiting its applicability in the clinic. 

Much more attractive are techniques for measuring SoS that do not require specialized equipment and can be implemented with handheld probes operating in standard pulse-echo B-mode \cite{robinson1991}. For example, SoS can be inferred by using strategies aimed at correcting aberration-induced image degradations \cite{hayashi1988,nock1989,shin2010,hesse2013,schiffner2014}. Recent strategies in this regard have been based on point-spread-function optimization making use of “guide stars” in the sample that are real or virtual \cite{odonnell1988,mallart1994,imbault2017,jakovljevic2018,byram2012}. Alternatively, matrix methods \cite{aubry2009} can be used to distinguish aberrations from sample using eigenmode decomposition \cite{lambert2019,bendjador2019}. 

Another class of techniques is based on analyzing the misregistration of images acquired with different transmit angles\cite{robinson1991,krucker2004}. This has led to the technique of computed ultrasound tomography in echo-mode (CUTE) \cite{jaeger2014}, where the misregistration is characterized by local phase shifts between different beamformed images (using a cross-correlation method as in \cite{loupas1995}). A ray-acoustic forward model linking these local phase shifts to SoS then allows the reconstruction of the SoS distribution in the sample using regularized inversion \cite{jaeger2014,jaeger2015,sanabria2018a,stahli2019}.

We present here an extension of the CUTE technique that provides differential phase contrast (DPC) with depth-dependent angular compounding to significantly improve contrast. An advantage of such depth-dependent angular compounding is that it helps reveal the rough locations of weakly aberrating inclusions without the need for numerical inversion. Like CUTE, our DPC technique is based on the now standard plane-wave compounding method of fast B-mode imaging \cite{montaldo2009}, with no changes whatsoever in hardware or data requirements. As such, our technique can be combined with B-mode imaging, providing a complementary contrast simultaneously and essentially for free. As a proof of concept, we adapted a commercial ultrasound machine to provide pulse-echo and DPC contrasts in elastography phantoms in both B and C modes. Moreover, we develop a forward model for our technique based on a 2D wave-acoustics formalism valid in the paraxial limit. We anticipate that such a forward model will be helpful in the future for more rigorous inversion-based SoS reconstruction.  

\section{DPC principle}

As noted above, our DPC strategy is based on the method of plane-wave compounding. In this method, which uses a standard linear-array transducer, the sample is insonified by a sequence of plane-wave pulses of differing tilt angles, producing a sequence of receive signals $\mathrm{RF}_n(x,t)$, where $x$ is the transverse actuator coordinate and  $n$ is the tilt-angle index. Upon beamforming, as described in \cite{montaldo2009}, these receive signals are converted to a sequence of images $B_n(x,z)$ where $z$ is the axial depth coordinate into the sample. Finally, the beamformed images $B_n(x,z)$ are coherently summed, in effect locally focusing the transmit pulses at each sample location.
 
It is well known that SoS variations within the sample can lead to local phase shifts in both the transmit plane waves and the receive spherical waves, causing aberrations in the beamformed images. The key premise in DPC (as in CUTE) is that the local phase shifts in the transmit waves are \emph{dependent} on the transmit tilt angle, whereas those in the receive waves are largely \emph{independent} of the transmit tilt angle. This premise is based on the beamforming assumption that the scatterers in the sample are point-like in nature (i.e. Rayleigh scatterers). In this manner, the scatterers themselves can serve as local probes of the transmit phase shifts. By monitoring, at every scatterer location, the changes in phase of $B_n(x,z)$ as a function of transmit tilt angle, one can in principle reconstruct the 2D distribution of SoS variations within the sample.
 
\begin{figure}[t]
	\centering
	\includegraphics[width=0.25\textwidth]{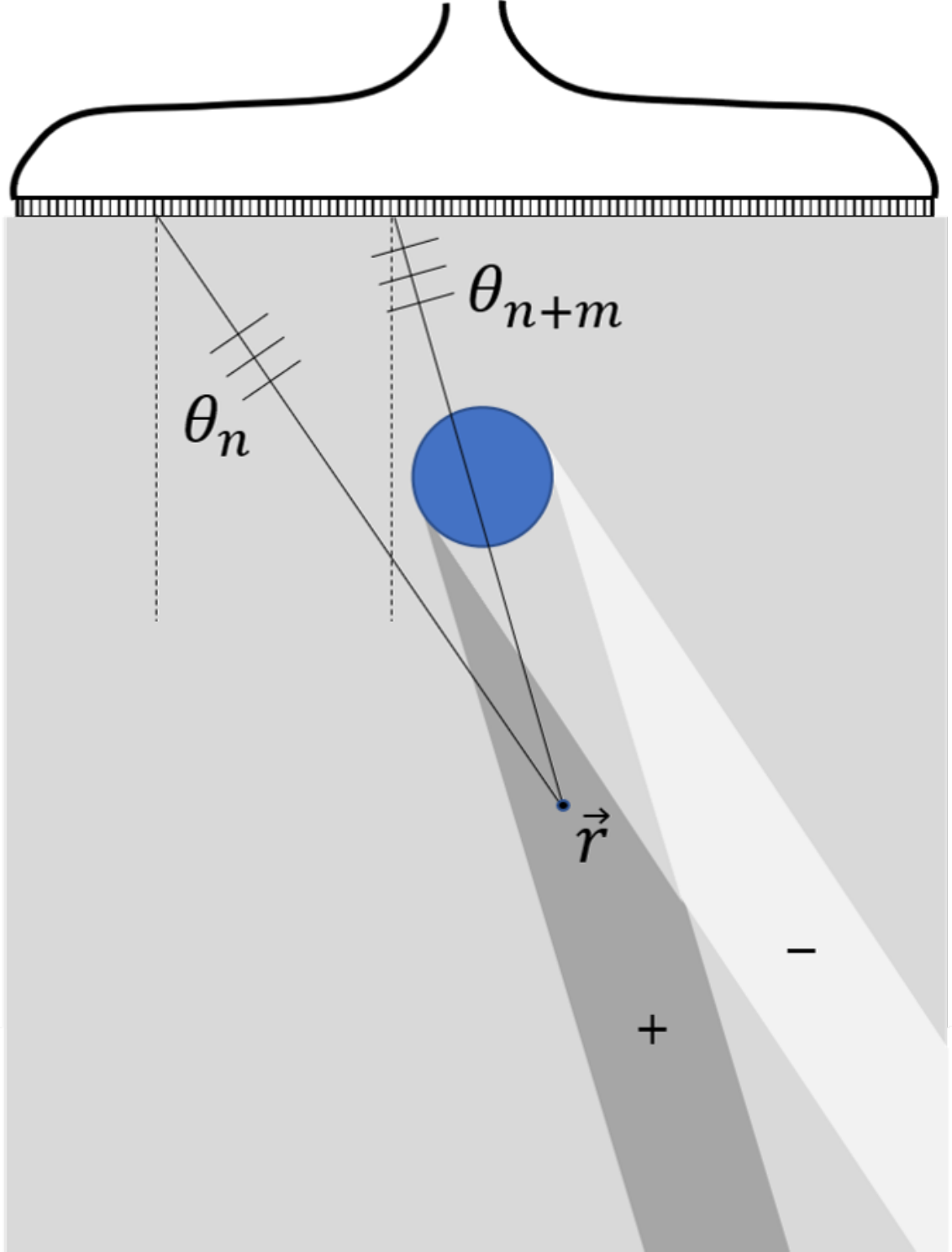}
	\caption{Two transmit plane-wave pulses of tilt angles $\theta_n$ and $\theta_{n+m}$ are incident on a spherical inclusion (blue) of lower SoS than its surrounding. Consider the evaluation of $\Delta\phi_n(\vec{r})$ at the indicated location $\vec{r}$. For this case, one pulse traverses the inclusion while the other does not, leading to a positive $\Delta\phi_n(\vec{r})$ (dark gray). Medium gray regions correspond to $\Delta\phi_n(\vec{r}) \approx 0$. Note that all transducer elements contribute to the generation of each plane wave pulse.   }
	\label{fig1}
\end{figure} 

\begin{figure}[h]
	\centering
	\includegraphics[width=0.3\textwidth]{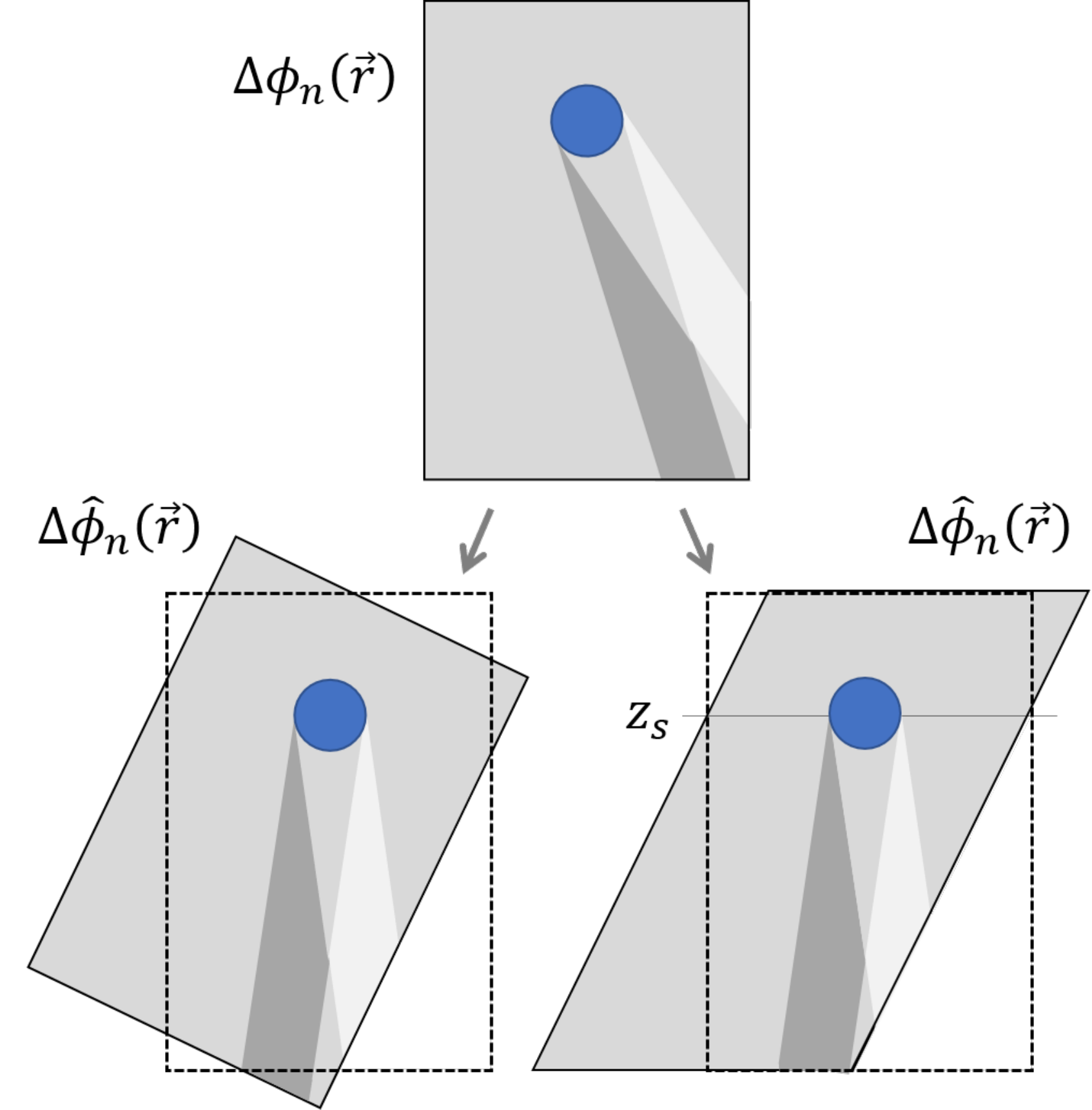}
	\caption{Illustration of two possible methods for untilting $\Delta\phi_n(\vec{r})$. Throughout this paper, we use the simpler method of image shearing about a user-defined depth $z_s$ (lower right).}
	\label{fig2}
\end{figure}
 
We note that the phase of $B_n(\vec{r})$ is in general a rapidly varying function of position (throughout this paper, we write $\vec{r}=\{x,z\}$ and consider $B_n(\vec{r})$ to be the complex analytical representation of the real beamformed image described in \cite{montaldo2009}). To compare the local phases of $B_n(\vec{r})$ and $B_{n+m}(\vec{r})$, where $n$ and $n+m$ correspond to two different transmit tilts, we simply subtract the phases. A robust method of doing this is to define the difference angle by $\Delta\phi_{n,m}(\vec{r})=\mathrm{arg}[B_n(\vec{r})B_{n+m}^*(\vec{r})]$, as described in \cite{jaeger2014} and akin to the phase of the angular coherence function described in \cite{li2017}. In principle, $m$ can be an arbitrary integer smaller than the total number of tilt angles $N$, however in this paper we mostly consider only nearest-neighbor tilt angles (i.e. $m=1$). In other words, we write $\Delta\phi_n(\vec{r})=\mathrm{arg}[B_n(\vec{r})B_{n+1}^*(\vec{r})]$, and suppress the index $m$.

Consider now the effect of a localized aberrator, such as a spherical inclusion of bulk modulus different than its surrounding (i.e. exhibiting different SoS), insonified sequentially by a pair of plane waves of tilt angles symmetrically distributed about the $z$-axis. From simple ray acoustics, it is clear that in the shadow regions downstream from the aberrators, $\Delta\phi_n(\vec{r})$ is either positive or negative depending on the sign of the SoS difference, except in the region of overlap where $\Delta\phi_n(\vec{r})$ is roughly zero (see Fig. \ref{fig1}).  From a careful analysis of $\Delta\phi_n(\vec{r})$, the nature of the aberrator can then, in principle, be reconstructed \cite{jaeger2014}.

However, there is a problem with this approach. In general the SoS variations in biological tissue, particularly soft tissue, are very small, leading to small values of $\Delta\phi_n(\vec{r})$ that exhibit poor contrast and are easily swamped by noise. We can of course make use of the fact that multiple pairs of tilt angles are at our disposal, but each of these creates shadow regions that are distributed about different midline tilt angles, and are thus pointing in different directions. A direct compounding (summation over $n$) of $\Delta\phi_n(\vec{r})$ tends to reduce contrast even further. On the other hand, if the midline tilt angles are numerically untilted prior to compounding so that they are all pointing in the vertical direction, then the contrast can be enhanced significantly, roughly by a factor $N-m$ (in our case $N-1$). We call this procedure angular compounding.

To perform the untilting process, each $\Delta\phi_n(\vec{r})$ should be rotated about the aberrator centroid, leading to $\Delta\hat{\phi}_n(\vec{r})$. However, the location of the aberrator centroid is unknown a priori. To explore all possible aberrator locations, we could apply this untilting procedure at every point $\vec{r}$ in the beamformed images, but this would be overly time consuming. Instead, we exploit the fact that the tilt angles utilized in plane wave compounding are generally small, meaning that an image rotation about point $\{x,z\}$ can be roughly approximated by an image shear about line $z$. In this manner, we do not have to explore all points $\{x,z\}$ to find the 2D location of the aberrator, but instead explore only $z$ to find the depth of the aberrator (see Fig. \ref{fig2}). This will be made clear below. In other words, our final DPC image depends of the depth one chooses for the shear line $z_s$, and is given by

\begin{equation}\label{PhiDPC}
\Delta\Phi(\vec{r},z_s)=\sum_{n=1}^{N-m} \Delta\hat{\phi}_n(\vec{r},z_s)
\end{equation}

\noindent
where $\Delta\hat{\phi}_n(\vec{r},z_s)$ is the untilted (by numerical shearing about $z_s$) version of $\Delta\phi_n(\vec{r})$, and $z_s$ can be adjusted at will. The ramifications of such adjusting will be discussed below.

\section{Forward model}\label{forward}

As will be shown in our experimental results, angular compounding
significantly improves the contrast of aberration-induced phase shifts. To
infer from this the actual distribution of aberrators within the sample, a key
first step is the development of a forward model that provides a quantitative
description of these phase shifts. This was done in \cite{jaeger2014} and \cite{stahli2019} using a 2D
ray-acoustic approach. Here we use a 2D wave-acoustic approach valid in the
paraxial (i.e. small-angle) limit. In particular, we consider an weak
aberrator located at an arbitrary depth in an otherwise homogeneous sample,
and derive an expression for $\Delta\Phi\left( \vec{r},z_{s}\right)  $ downstream
from this depth, assuming that $z_{s}$ has been preassigned to the aberrator
depth. The aberrator is modeled as a thin structure of spatially varying SoS.
In other words, planes-wave pulses traveling through the aberrator incur
spatially varying time delays (or advances), defined here by $\tau_{a}\left(
x\right)  $, where $\tau_{a}\left(  x\right)  =0$ wherever the aberrator is
absent. For ease of notation, we adjust our coordinate system such that the
depth of the aberrator is set to $z=z_{s}=0$. A full derivation of $\Delta
\Phi\left( \vec{r},z_{s}=0\right)  $ is provided in the Appendix. We provide here
only the final result, given by

\begin{align}\label{PhiDPCforward}
\begin{split}
&\Delta\Phi\left(  \vec{r}\right)  = \arg \Bigg[  \int W_{\varphi}\left(  \vec{r}%
,\tau_{a},x_{c}\right)  \\
&  \times  \exp\left[-\pi^{2}\theta_{d}^{2}\kappa_{\sigma}^{2}\left( %
x-x_{c}\right)  ^{2}-i2\pi\theta_{d}\kappa_{0}\left(  x-x_{c}\right) \right] %
dx_{c} \Bigg]
\end{split}
\end{align}

\noindent where

\begin{equation}\label{W}
W_{\varphi}\left(  \vec{r},\tau_{a},x_{c}\right)  =\frac{1}{\varphi_{\kappa}%
}\exp\left[  -\frac{\pi^{2}}{\varphi_{\kappa}^{2}}\left(  \frac{x-x_{c}}%
{z}-c\frac{d}{dx}\tau_{a}\left(  x_{c}\right)  \right)  ^{2}\right] 
\end{equation}

\begin{figure}[h]
	\centering
	\includegraphics[width=0.31\textwidth]{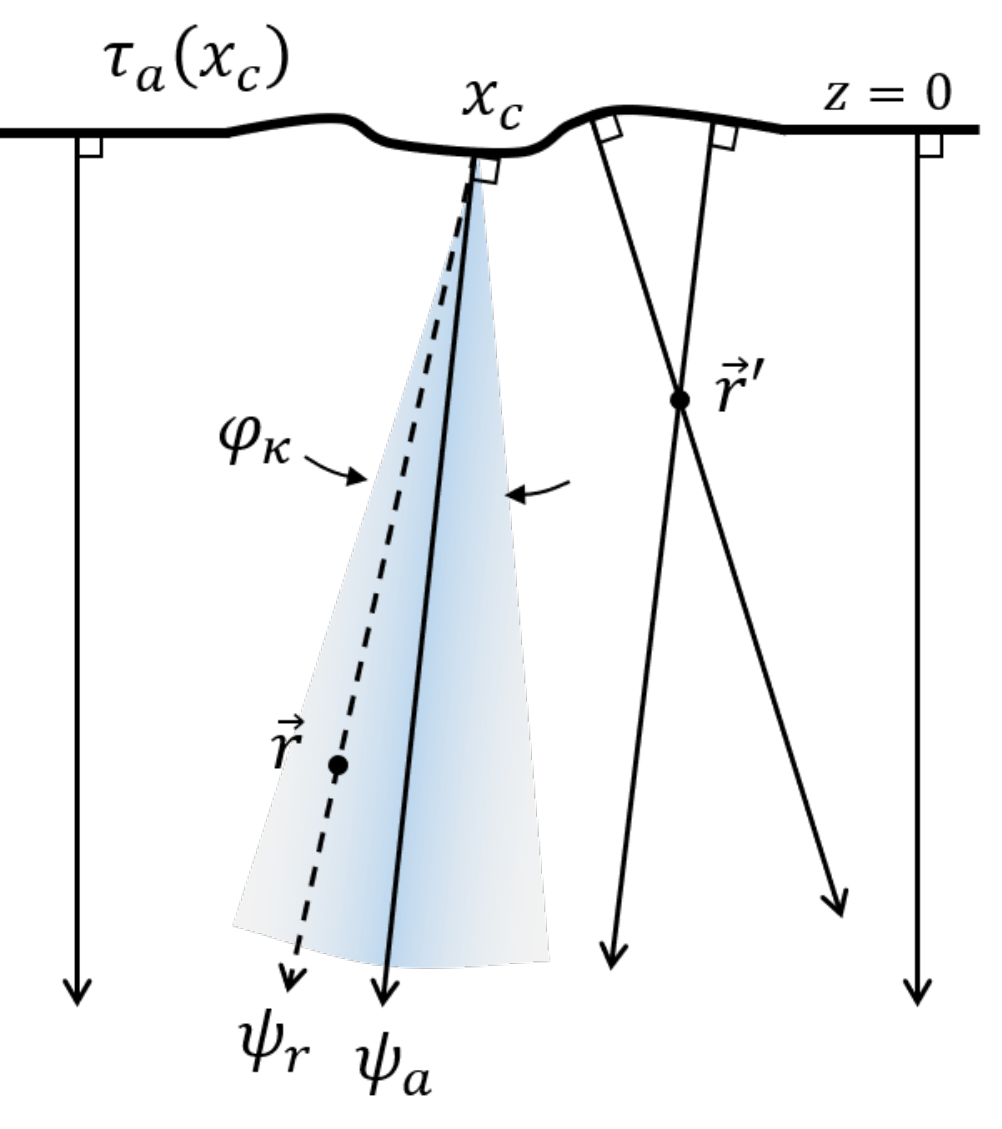}
	\caption{Geometry associated with forward model. A thin aberrator located at $z=0$ produces spatially varying time delays $\tau_{a}(x_{c})$. These define a ray $\psi_a$ emanating from each location $x_c$ was well as an associated weighting function $W_{\varphi}\left(  \vec{r},\tau_{a},x_{c}\right)$ about $\psi_a$ (in blue). Some locations in the sample (e.g. $\vec{r}^{\,\prime}$) are intersected by rays emanating from multiple $x_c$'s.    }
	\label{fig3}
\end{figure}

In the above, $\vec{r}$ is an arbitrary downstream
location in the sample, $\kappa_{0}$ and $\kappa_{\sigma}$ are respectively the wavenumber
average and wavenumber spread of the plane-wave pulses, $c$ is
the background SoS, $\theta_{d}$ is the angular separation between
pairs of incident plane-wave pulse directions used to produce DPC, and
$\varphi_{\kappa}$ is a phenomenological coherence parameter associated with
the incident plane-wave pulses. The greater the transverse spatial coherence
of the pulses, the smaller $\varphi_{\kappa}$ (more on this below).

The interpretation of $W_{\varphi}\left(  \vec{r},\tau_{a},x_{c}\right)  $ is
straightforward. Consider a particular transverse location $x_{c}$ at the
aberrator depth ($z=0$), as shown in Fig. \ref{fig3}. Two angular directions may be
defined. The first is $\psi_{r}\left(  x_{c}\right)  =\left(  x-x_{c}\right)
/z$ characterizing the direction of a ray from $x_{c}$ to $\vec{r}$. The
second is $\psi_{a}\left(  x_{c}\right)  =c\frac{d}{dx}\tau_{a}\left(
x_{c}\right)  $ characterizing the direction of a ray normal to the slope of
$c\tau_{a}\left(  x_{c}\right)  $. In this manner, we can recast Eq. \ref{W} as

\begin{equation}\label{W1}
W_{\varphi}\left(  \vec{r},\tau_{a},x_{c}\right)  =\frac{1}{\varphi_{\kappa}%
}\exp\left[  -\frac{\pi^{2}}{\varphi_{\kappa}^{2}}\left(  \psi_{r}\left(
x_{c}\right)  -\psi_{a}\left(  x_{c}\right)  \right)  ^{2}\right] 
\end{equation}

\noindent revealing that $W_{\varphi}\left(  \vec{r},\tau_{a},x_{c}\right)  $
plays the role of a weighting function. The greater the angular deviation
between $\psi_{r}\left(  x_{c}\right)  $ and $\psi_{a}\left(  x_{c}\right)  $,
the more attenuated $W_{\varphi}\left(  \vec{r},\tau_{a},x_{c}\right)  $
becomes, where the range of angular deviations allowed by $W_{\varphi}\left(
\vec{r},\tau_{a},x_{c}\right)  $ is determined by $\varphi_{\kappa}$.
Indeed, when $\varphi_{\kappa}\rightarrow0$ then, roughly, $W_{\varphi}\left(
\vec{r},\tau_{a},x_{c}\right)  \rightarrow\delta\left(  \psi_{r}\left(
x_{c}\right)  -\psi_{a}\left(  x_{c}\right)  \right)  $ and no deviation is
allowed at all. In this case, our wave-acoustic model reduces to a
ray-acoustic model.

\begin{figure}[h!!]
	\centering
	\includegraphics[width=0.35\textwidth]{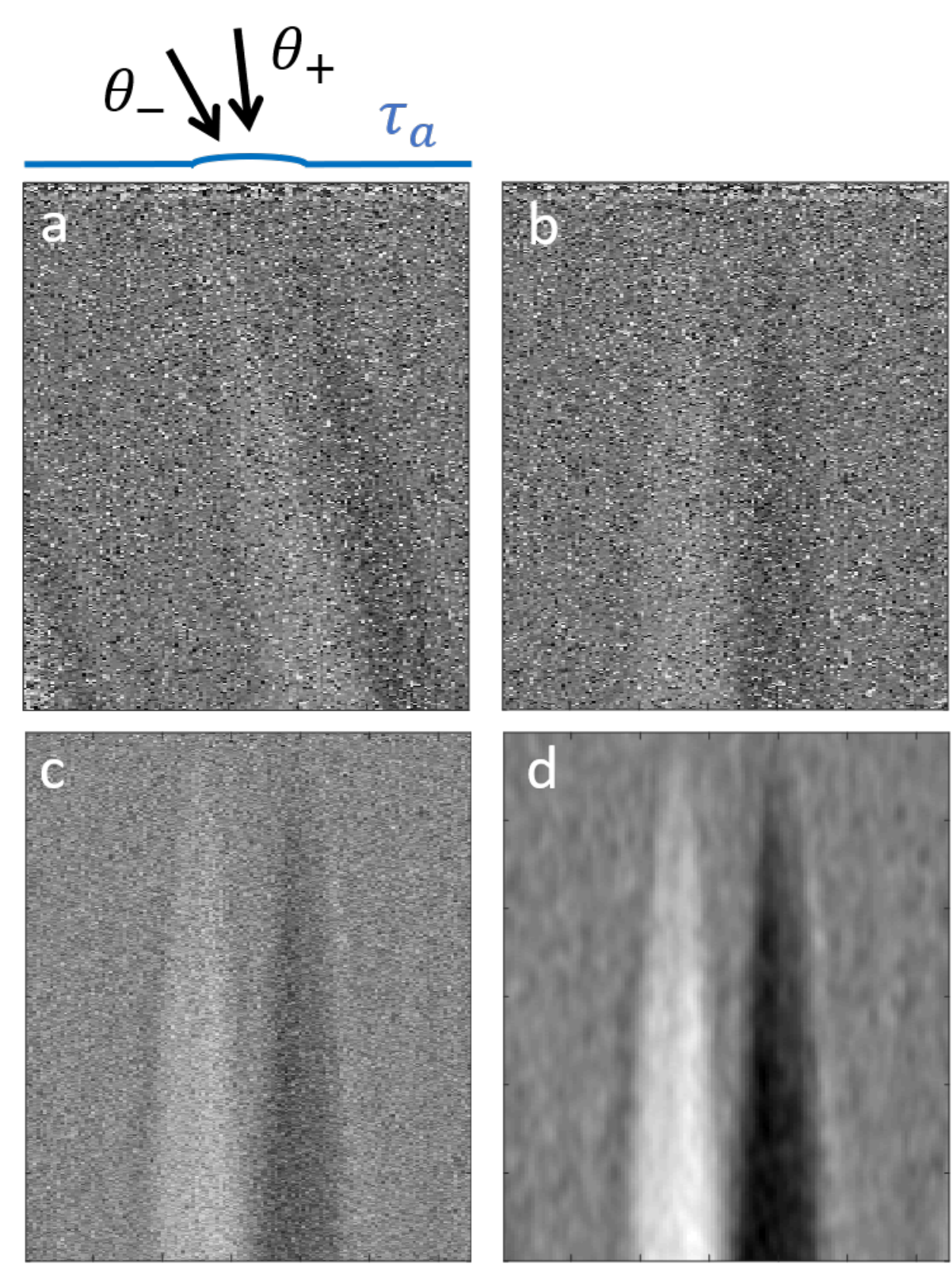}
	\caption{Experimental results with simulated aberration. A time delay $\tau_a(x_0)$ was added to all transmit pulses to simulate a spherical aberration of radius $16\lambda_0$ and maximum time delay $0.2\lambda_0/c$, where $\lambda_0=0.3$mm is the center wavelength. A DPC image obtained from the first pair of nearest-neighbor angles from a 7-angle transmit sequence (a) before untilting, (b) after untilting (shear depth set to $z=0$), (c) after angular compounding over all transmit angles, and (d) after Gaussian filtering. Here, $\theta_d=\theta_{+}-\theta_{-}=0.06$. Panel sizes $150\lambda_0\times 150\lambda_0$ (aspect ratio is not unity).}
	\label{fig4}
\end{figure}

\begin{figure*}[t!]
	\centering
	\includegraphics[width=0.9\textwidth]{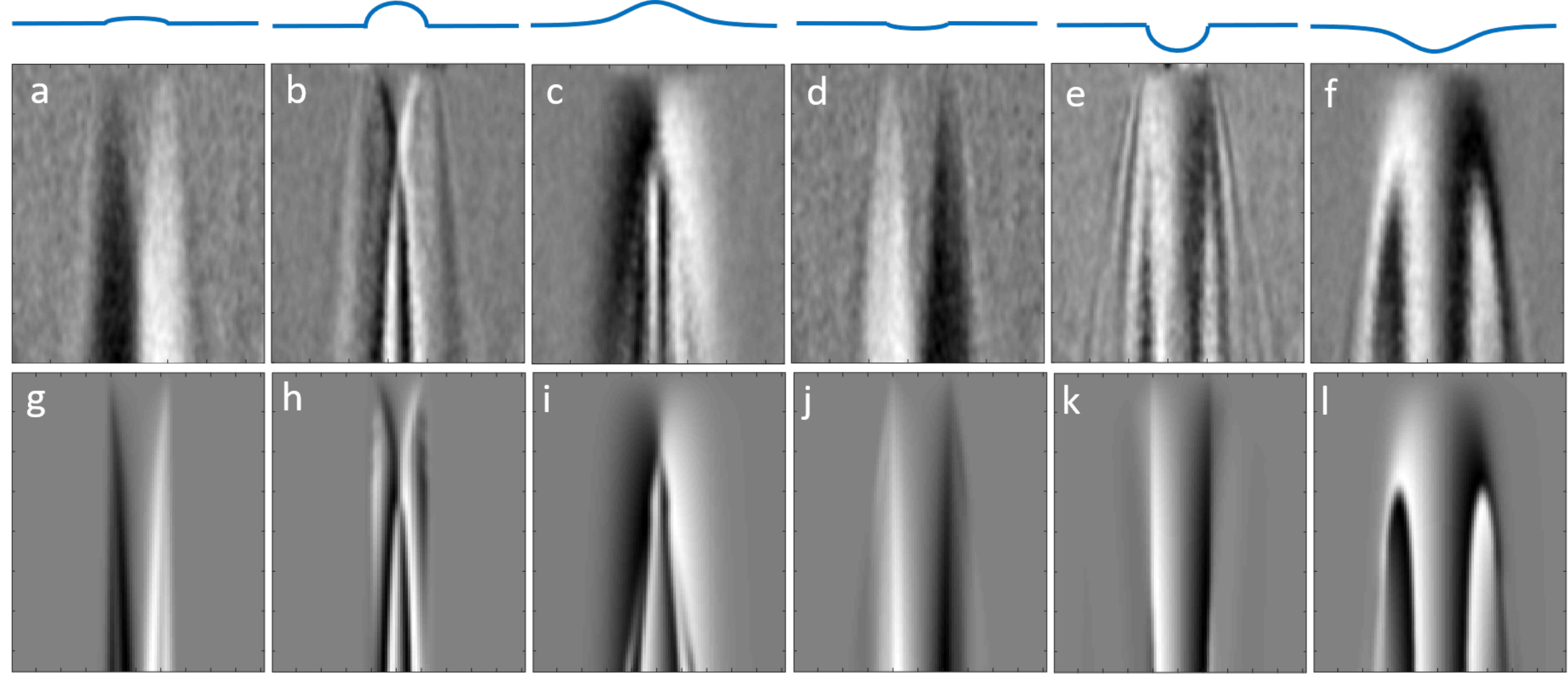}
	\caption{Comparison of experimental (top row) and theoretical (bottom row) results for different simulated aberrations. Simulations of spherical aberrations of radius $16\lambda_0$ and maximum time delay (a,g) $0.2\lambda_0/c$, (d,j) $-0.2\lambda_0/c$, (b,h) $2\lambda_0/c$, and (e,k) $-2\lambda_0/c$. Simulations of Gaussian  aberrations of waist $16\lambda_0$ and maximum time delay (c,i) $2\lambda_0/c$, and (f,l) $-2\lambda_0/c$. Panel sizes $150\lambda_0\times 150\lambda_0$.     }
	\label{fig5}
\end{figure*}

With this interpretation of $W_{\varphi}\left(  \vec{r},\tau_{a},x_{c}\right)
$, we can now better understand Eq. \ref{PhiDPCforward}, at least in the ray-acoustic limit.
Consider an arbitrary location $\vec{r}$ in the sample. One or many rays
$\psi_{a}\left(  x_{c}\right)  $ roughly intersect this location. For a
particular harmonic component of wavenumber $\kappa$, the corresponding DPC
signal is given by the coherent summation of $\exp\left[  -i2\pi\theta
_{d}\kappa\left(  x-x_{c}\right)  \right]  $ over all these intersecting rays, where
$\left(  x-x_{c}\right)$ is the transverse separation between the ray location expected by beamforming
and the actual ray location deviated by the aberrator. 
Finally, the integration over all
harmonic components encompassed by the transducer bandwidth leads to the full DPC signal.
This is made clear in the Appendix.

\section{Experimental results}

To validate our method of DPC as well as our wave-acoustic forward model, we used a Verasonics Vantage 256 to perform conventional plane-wave compounding. Our transducer was a GE9L-D linear array (192 elements, 0.23mm pitch) of center frequency 5.3 MHz, and our sampling rate was adjusted to be $4\times$ the center frequency. For our experimental results, we made use of a CIRS elasticity QA phantom (model 049) in which was embedded a variety of spherical inclusions of Young’s moduli differing from the background. For our initial results, we made use of our own beamformer; for later results, we integrated DPC directly into the Verasonics software and made use of the beamformer provided by Verasonics.

\begin{figure}[h]
	\centering
	\includegraphics[width=0.35\textwidth]{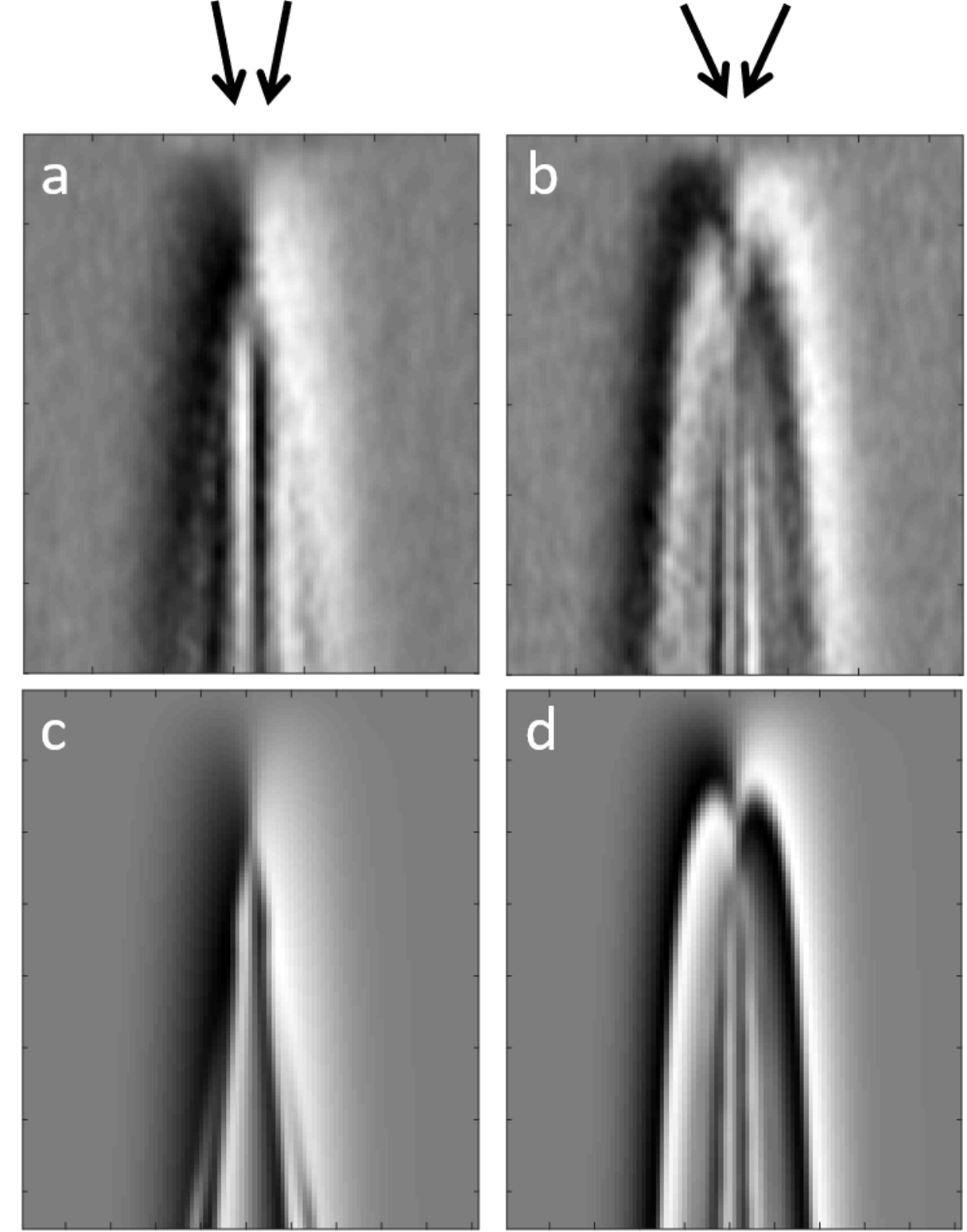}
	\caption{Comparison of experimental (top row) and theoretical (bottom row) results for simulated Gaussian aberrations (same as Fig. \ref{fig5}(c,i)), with different tilt angle separations. (a,c) $\theta_d=0.06$, (b,d) $\theta_d=0.12$. Panel sizes $150\lambda_0\times 150\lambda_0$.}
	\label{fig6}
\end{figure}

To begin, we did not image the spherical inclusions in our phantom. Instead, we directed our transducer over a region known to be featureless. This enabled us to introduce our own simulated aberrations by simply adding user-defined time delay functions to the transmit pulses. In other words, this enabled us to control $\tau_{a}(x_{c})$ directly. For example, in Fig. \ref{fig4} we adjusted $\tau_{a}(x_{c})$ to mimic a spherical inclusion. We insonifed the sample with 7 plane-wave tilt angles in total, spanning -12 to 12 degrees (the same $\tau_{a}(x_{c})$ was applied to each tilt angle). Fig. \ref{fig4} illustrates the effectiveness of angular compounding at dramatically increasing DPC contrast. Note that the shear depth here was set to $z_s = 0$.

\begin{figure}[h]
	\centering
	\includegraphics[width=0.35\textwidth]{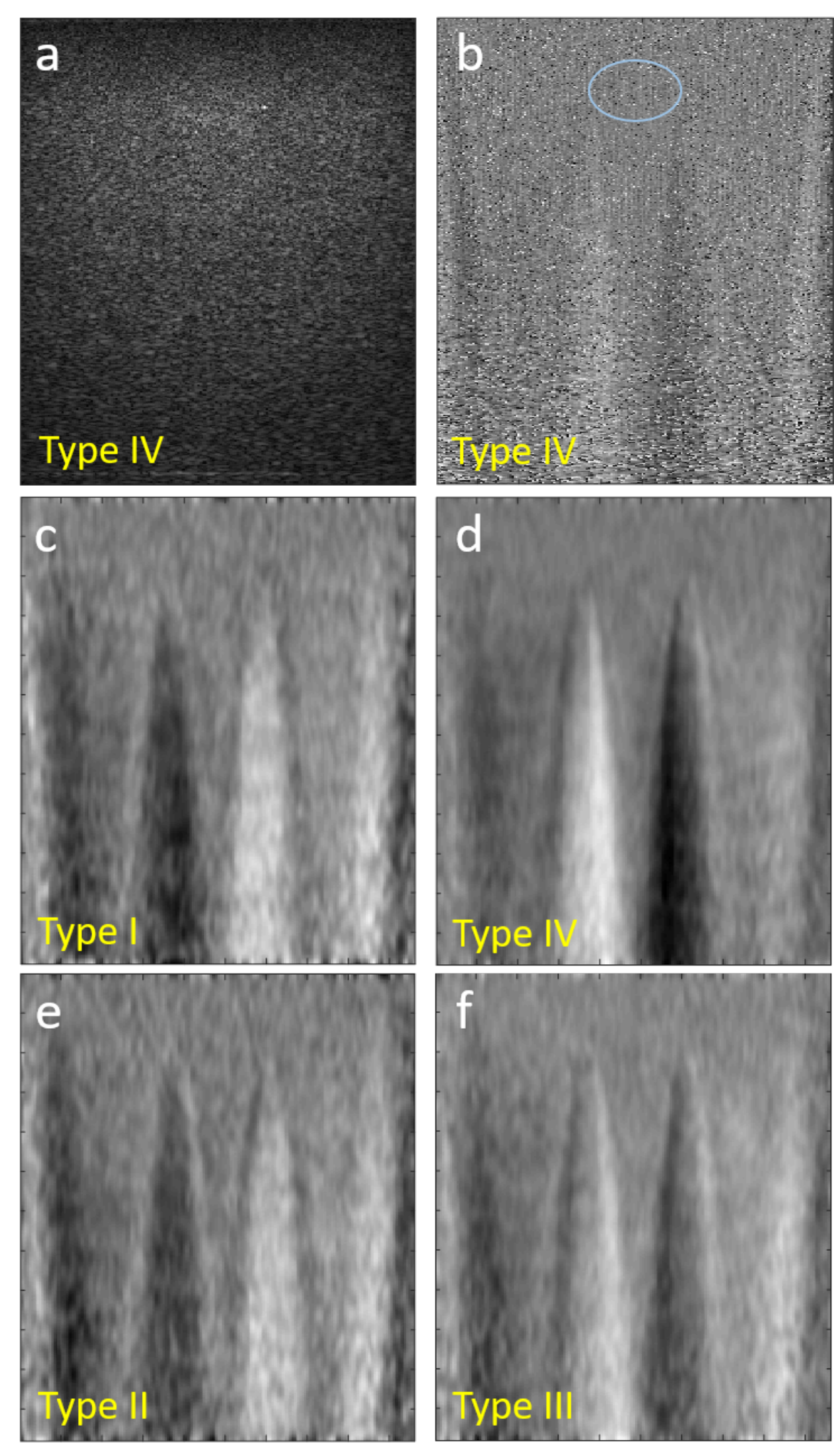}
	\caption{Images of CIRS elasticity phantom obtained with Verasonics Vantage. (a) Conventional pulse-echo image of Type IV inclusion obtained by plane-wave compounding (15 transmit angles). (b) Single DPC image from central pair of transmit angles (outline of inclusion shown in blue). (c-f) Angular-compounded DPC images for all inclusion types. Shear depths were set to inclusion depths, here $z=15$mm. Note that pulse-echo and DPC images were created from the same raw data. Panel sizes $150\lambda_0\times 300\lambda_0$. }
	\label{fig7}
\end{figure}

To compare experimental results with those predicted by our forward model, we varied $\tau_{a}(x_{c})$ to mimic different shaped inclusions of differing SoS variations (Fig. \ref{fig5}). As expected, changing the sign of $\tau_{a}(x_{c})$ (corresponding to a faster or slower SoS than the surrounding), roughly led to an overall change of sign in the DPC contrast. However, our forward model also takes into account both diffraction and refraction effects, leading to DPC images that not only change in sign, but are qualitatively different in appearance. For a ray-acoustic model to reproduce such effects, it would have to take into account the bending of rays, either inward or outward, caused by the inclusions (e.g. \cite{clement2003}). In general, our forward model accurately captures the most salient features in our DPC images, particularly for small angles and in regions directly downstream from the aberration. For larger angles that extend beyond these regions, our model is not as accurate, for reasons that are still unclear. For example, our model does not account for the diffraction effects observed in panel \ref{fig5}(e), which likely result from the fact that our transducer actuators are discrete and not continuous.     

\begin{figure}[h]
	\centering
	\includegraphics[width=0.35\textwidth]{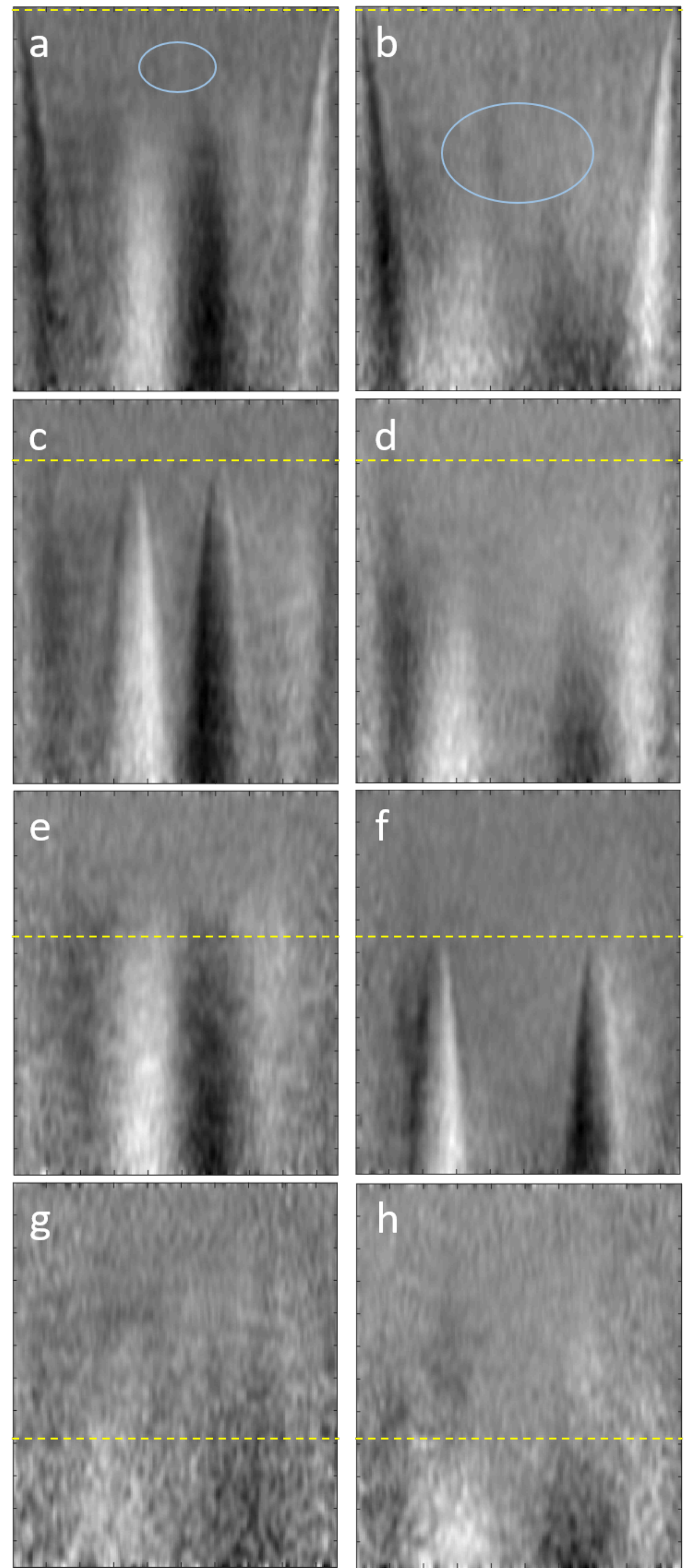}
	\caption{DPC images of two different Type IV inclusions located at different depths (left column: smaller inclusion at 15 mm depth; right column: larger inclusion at 35 mm depth - outlines shown in (a,b)). Angular compounding was performed with shear depths (yellow dashed) set to (a,b) $z=0$mm, (c,d) $z=15$mm, (e,f) $z=35mm$, and (g,h) $z=60$mm. Panel sizes $150\lambda_0\times 300\lambda_0$.  }
	\label{fig8}
\end{figure}

The results shown in Figs. \ref{fig4} and \ref{fig5} were obtained using nearest-neighbor tilt angles. That is, $m=1$ and $\theta_d = 0.06$. However, our forward model is not limited to nearest-neighbor tilt angles, and indeed generalizes to arbitrary (albeit small) tilt angle separations. Fig. \ref{fig6} shows the result of using, for example, $m=2$ in our DPC processing algorithm (using the same raw data). Here, too, our forward model captures the salient features in our experimental results.

We turn now to DPC imaging of the spherical inclusions themselves. To perform this, we integrated our DPC algorithm directly into the Verasonics software, enabling us to obtain and display standard pulse-echo and DPC images essentially simultaneously, using the Verasonics beamformer for both (with 15 tilt angles, our Intel Xeon 2.5 GHz computer displayed pulse-echo images at a frame rate of about 5 Hz; with the addition of DPC display this frame rate dropped to about 2.5 Hz).

\begin{figure}[h]
	\centering
	\includegraphics[width=0.35\textwidth]{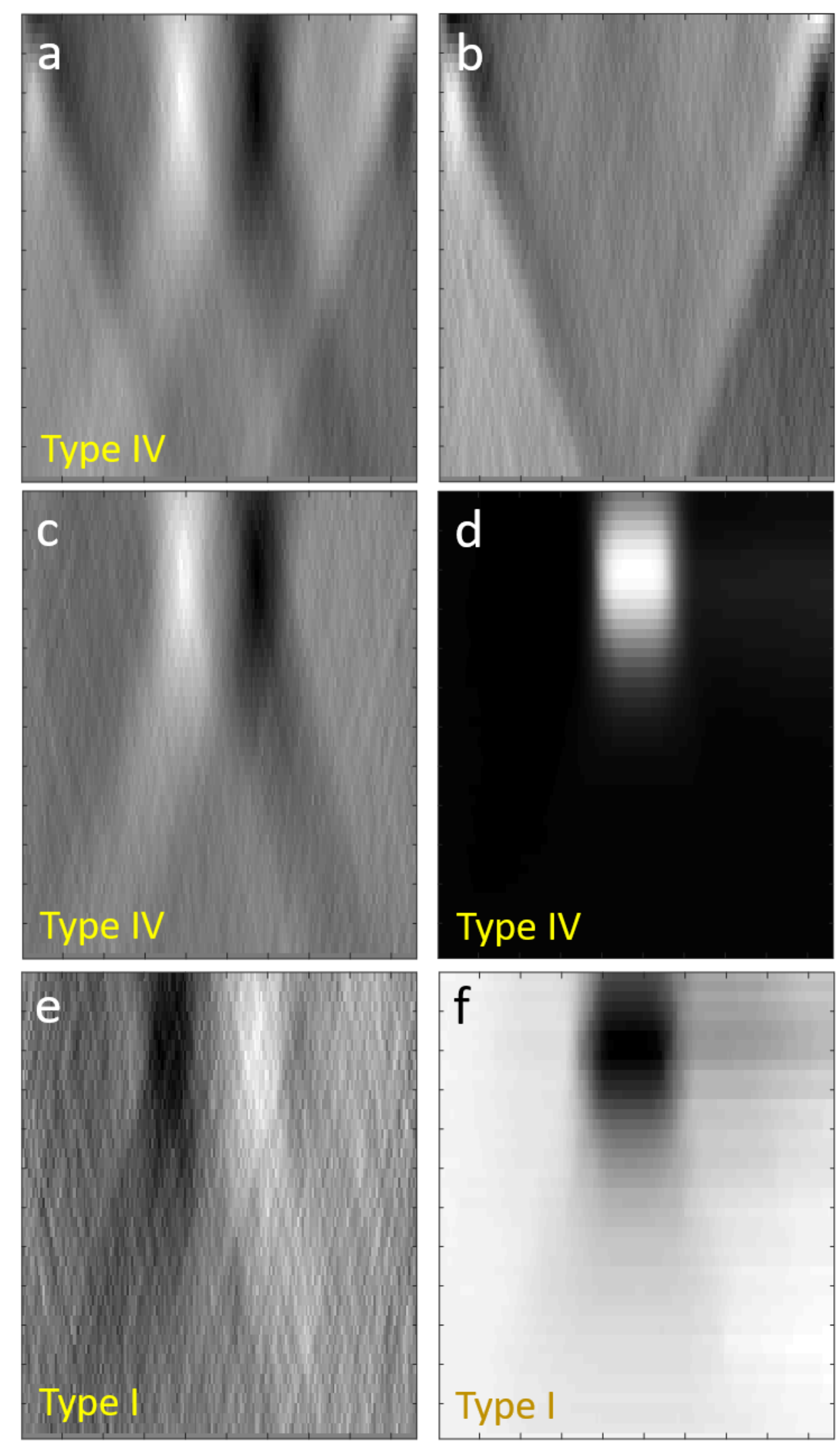}
	\caption{``In-focus" DPC images of Type IV and I inclusions. Each row is the extended-field projection (integration over $z$) of the DPC image obtained for different shear depths (vertical axis). Panel (c) is same as (a) but corrected for edge artifacts (b). Panels (d,f) are the images obtained by transverse integration of (c,e) after nonlinear enhancement (pixel values in (c,e) raised to the 3rd power prior ro integration, to arbitrarily sharpen edges). Panel sizes $150\lambda_0\times 300\lambda_0$. }
	\label{fig9}
\end{figure}

\begin{figure}[h!]
	\centering
	\includegraphics[width=0.35\textwidth]{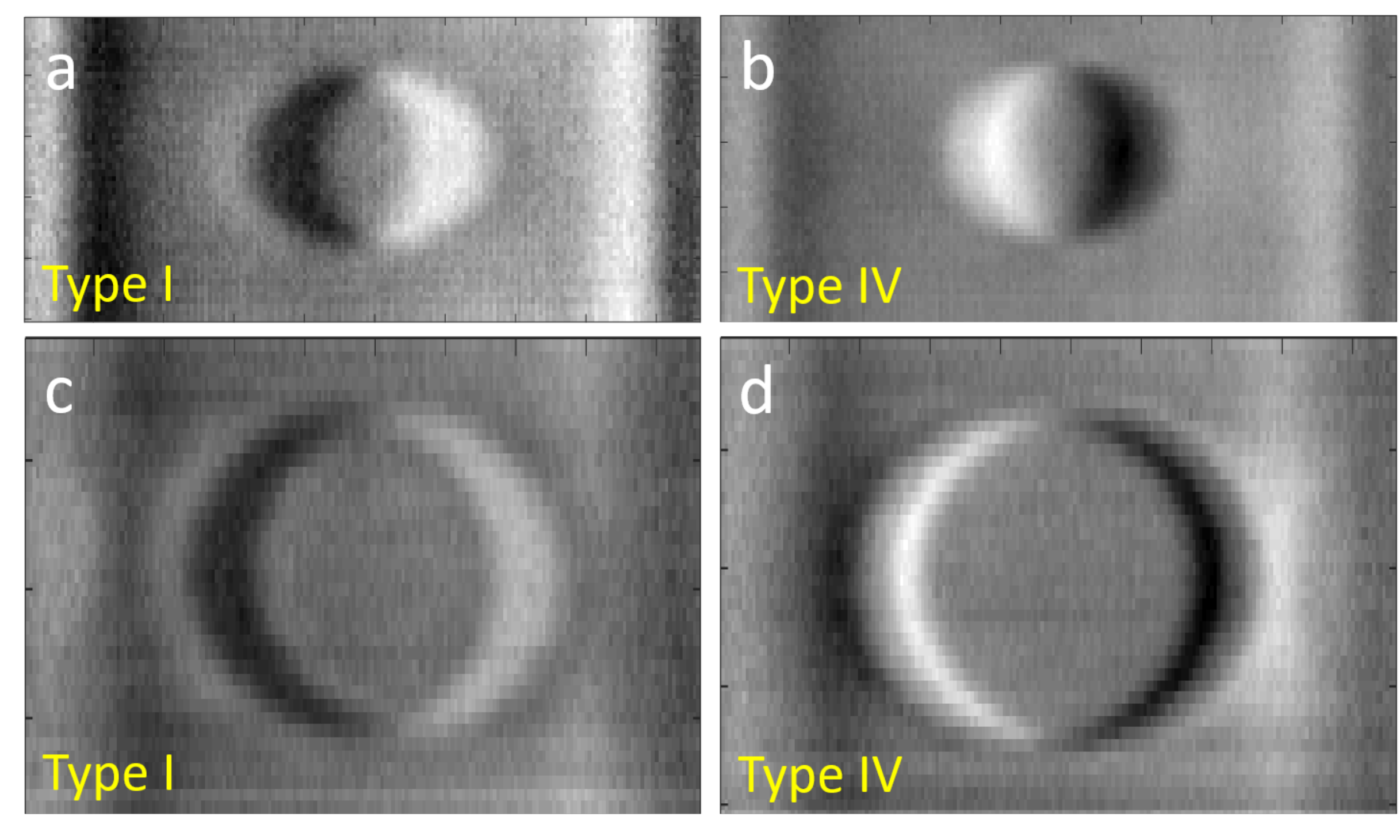}
	\caption{C-mode images of Type I and IV inclusions. (a,b) inclusions of smaller sizes and depths; (c,d) inclusions of larger sizes and depths. C-mode images obtained by translating transducer in $y$-direction (vertical) and integrating DPC signals over all depths. Note reversal of contrast depending on inclusion type (relative decrease or increase of  SoS). Horizontal panel sizes $150\lambda_0$. }
	\label{fig10}
\end{figure}

Our phantom contains four different types of inclusions, labeled Type I-IV, of Young’s moduli 6 kPa, 9 kPa, 36 kPa and 70 kPa respectively (the background modulus is 18 kPa), of different sizes and depths. In Fig. \ref{fig7} we show images of the smaller inclusions (diameter 10 mm) at shallower depths (15 mm). We observe that while the inclusions are hardly visible in pulse-echo mode (Fig. \ref{fig7}a), their signatures are quite apparent in DPC mode. We note that the expected SoS within these inclusions is respectively $1530\pm15$ m/s, $1533\pm15$ m/s, $1552\pm15$ m/s and $1572\pm10$ m/s, with background SoS $1540\pm15$ m/s (based on manufacturer specifications), leading to time-delay-induced phase shifts of fractions of a wavelength.

To obtain the DPC images in Fig. \ref{fig7} we set the shear depth to be the same as the known depth of the inclusions (i.e. $z_s=15$ mm). Figure \ref{fig8} shows the effect of varying this shear depth arbitrarily. The top three panels are images of a smaller, shallower inclusion; the bottom three panels are images of a larger, deeper inclusion (diameter 20 mm; depth 35 mm). We emphasize that $z_s$ in all cases was varied post-acquisition using the same raw data, and was controlled in real time using a simple slider added to our Verasonics GUI. As is apparent from Fig. \ref{fig8}, the correct adjustment  of $z_s$ plays a key role in maximizing DPC contrast, and indeed, $z_s$ can be thought of as a kind of “focus knob” enabling us to estimate the depth of the inclusion by qualitatively maximizing DPC sharpness.

Figure \ref{fig9} provides an illustration of this focusing ability. Here, the vertical axis corresponds to the location of the shear plane, which was numerically scanned from the sample surface to the maximum imaging depth. For each shear-plane depth, the resultant $x$-$z$ DPC image was integrated over all $z$’s, thus producing the image row at the shear-plane depth. As can be observed in Fig. \ref{fig9}a, the approximate location and depth of the inclusion is readily apparent. We note that we made no attempt here to correct for the edge artifacts arising from the incomplete coverage of the sample with our tilted transmit beams (nor did we make any such attempt in previous figures). However, these artifacts can be assessed by performing DPC in a region of the sample that is known to be free of inclusions (Fig. \ref{fig9}b), and subsequently subtracted out (Figs. \ref{fig9}c,e). These results highlight the capacity of DPC to reveal not the SoS variations themselves, but rather their transverse gradient (similar to the analog of DPC in x-ray \cite{david2002} or optical imaging \cite{mehta2009}). For example, a very rough, non-quantitative reconstruction of the inclusion can be inferred by simple transverse integration, where, to better highlight the inclusion edges, we artificially amplified the gradient nonlinearly (Figs. \ref{fig9}d,f).

Finally, we adapted our machine to perform C-mode imaging. For this, we mounted our transducer onto a motorized translation stage (Thorlabs MTS50-Z8) which translated the transducer in the $y$-direction with a manually controlled speed (Thorlabs KDC101). Figure \ref{fig9} shows extended depth of field DPC images. That is, for each $y$ position, the $x$-$z$ DPC image was integrated over all $z$’s, similarly to above but with the shear depth set to the inclusion depth. The resultant $x$-lines from approximately 25 $y$ positions were then assembled into $x$-$y$ images (Fig. \ref{fig10}  -- without edge artifact corrections). In our software, this assembling of $x$-$y$ images was performed on the fly and displayed in a scrolling window. What is apparent from these images is that the inclusions are easily recognizable as spheres. Moreover, the sign of the SoS variations (or equivalently the sign of the Young’s modulus deviation from background) could be immediately inferred from the relative orientation of the dark versus bright regions about the inclusion edges.

\section{Conclusion}
In summary, we have presented an angular compounding strategy that quickly and simply reveals transmit phase shifts caused by aberration-induced SoS variations with high contrast. The strategy can be implemented in parallel with standard pulse-echo imaging based on plane-wave compounding, and displays both pulse-echo and phase-shift contrasts simultaneously. We emphasize that we do not yet perform actual SoS reconstructions of our samples, as is done for example with CUTE \cite{jaeger2014,stahli2019}. Instead, our DPC method is meant simply to help identify the presence of aberrating inclusions in real time, along with some of their qualitative features (sign of SoS variation, rough 3D location, etc.). Ultimately, we hope to complement DPC with an algorithm for full SoS reconstruction, and for this we developed a forward model to describe DPC. At present, our wave-optic formalism considers only a single aberration layer, which is treated as a boundary condition when solving for 2D wave propagation in an otherwise homogeneous medium. Extension of this formalism to 3D is straightforward, using a 3D instead of 2D paraxial (Fresnel) propagator. Extension beyond the paraxial approximation is also potentially straightforward using the Rayleigh-Sommerfeld propagator \cite{goodman2017}, though less amenable to analytic integration. Finally, extension to the modeling of multiplane or volumetric aberrations could be implemented using formalisms borrowed from the optics community, such as the beam propagation method \cite{van1981} or Rytov formalism \cite{rytov1988}.              

\section*{Appendix}

We provide here a derivation of Eq. \ref{PhiDPCforward}, which will be separated into two parts.
In the first part we derive a general expression for the DPC signal. In the
second part, we consider the effect of an aberrator.

\textit{DPC signal}

To begin, we write the general equation for the construction of a beamformed
image \cite{montaldo2009} obtained with a single plane-wave pulse. In $\{\vec{r},\nu\}$ space,
this is given by

\begin{align}\label{Bn}
\begin{split}
B_{n}\left(  \vec{r}\right)  &=\int d\nu\iint e^{-i2\pi\nu\left(  \tau
	_{r}\left(  \vec{r},\vec{r}_{2}\right)  +\tau_{e}\left(  \vec{r},\hat{s}
	_{n}\right)  \right)  } \\
&\times  G\left(  \vec{r}_{2}-\vec{r}_{1},\nu\right)
\gamma\left(  \vec{r}_{1}\right)  P\left(  \vec{r}_{1},\nu,\hat{s}_{n}\right)
d\vec{r}_{2}d\vec{r}_{1}
\end{split}
\end{align}

\noindent where $\vec{r}$ and $\vec{r}_{1}$ are sample coordinates, $\vec
{r}_{2}=\left\{  x_{2},0\right\}  $ is the transducer coordinate, and
$P\left(  \vec{r}_{1},\nu,\hat{s}_{n}\right)  =P_{0}\exp\left[ -\left(  \nu-\nu
	_{0}\right)  ^{2}/2\nu_{\sigma}^{2}+i2\pi\nu\hat{s}_{n}\cdot\vec{r}_{1}/c \right]$
is the incident plane-wave pulse of center frequency $\nu_{0}$, frequency
bandwidth $\nu_{\sigma}$ (assumed Gaussian), and propagation direction
$\hat{s}_{n}$ (a unit vector). The sample is assumed to be comprised of
point-like scatterers of reflectivity distribution $\gamma\left(  \vec{r}%
_{1}\right)  $, $G\left(  \vec{r}_{2}-\vec{r}_{1},\nu\right)  $ is the Green
function transporting sample reflections back to the transducer to be
received, and the exponential in the integrand corresponds to the process of
beamforming itself, where $\tau_{e}\left(  \vec{r},\hat{s}_{n}\right)
=\hat{s}_{n}\cdot\vec{r}/c$ and $\tau_{r}\left(  \vec{r},\vec{r}_{2}\right)
=\left\vert \vec{r}-\vec{r}_{2}\right\vert /c$ are the transmit and receive
time-delay corrections associated with sample position $\vec{r}$.

Henceforth, we consider a monochromatic wave and suppress $\nu$ in our
notation (the integration over $\nu$ will be relegated to the end of our
calculation). We write then

\begin{equation}\label{Bn1}
B_{n}\left(  \vec{r}\right)  _{\nu}=e^{-i2\pi\nu\tau_{e}\left(  \vec{r}%
	,\hat{s}_{n}\right)  }\int H_{r}\left(  \vec{r}-\vec{r}_{1}\right)
\gamma\left(  \vec{r}_{1}\right)  P\left(  \vec{r}_{1},\hat{s}_{n}\right)
d\vec{r}_{1}
\end{equation}

\noindent where we have defined the receive amplitude point-spread function to be

\begin{equation}\label{H}
H_{r}\left(  \vec{r}-\vec{r}_{1}\right)  =\int e^{-i2\pi\nu\tau_{r}\left(
	\vec{r},\vec{r}_{2}\right)  }G\left(  \vec{r}_{2}-\vec{r}_{1}\right)  d\vec
{r}_{2}
\end{equation}

We recall that to obtain a DPC image, we must interfere (multiply) pairs of
beamformed images obtained from different plane-wave transmit directions. That
is, we must evaluate $B_{+}\left(  \vec{r}\right)  B_{-}^{\ast}\left(  \vec
{r}\right)  _{\nu}$, where $\hat{s}_{\pm}$ denote two transmit directions.

At this point, we can make an assumption about our sample, namely that the scatterers are
\emph{randomly} distributed throughout the sample in a roughly homogeneous
manner. That is, we assume $\left\langle \gamma\left(  \vec{r}_{1}\right)
\gamma\left(  \vec{r}_{1}^{\,\prime}\right)  \right\rangle \approx\bar{\gamma
}\delta\left(  \vec{r}_{1}-\vec{r}_{1}^{\,\prime}\right)  $, where the
brackets indicate local spatial averaging. We readily find then

\begin{align}\label{DPC1}
\begin{split}
B_{+}\left(  \vec{r}\right) & B_{-}^{\ast}\left(  \vec{r}\right)  _{\nu}%
=\bar{\gamma}e^{-i2\pi\kappa\left(  \hat{s}_{+}-\hat{s}_{-}\right)  \cdot
	\vec{r}} \\
&\times  \int \mathrm{PSF}_{r}\left(  \vec{r}-\vec{r}_{1}\right)  P\left(  \vec{r}%
_{1},\hat{s}_{+}\right)  P^{\ast}\left(  \vec{r}_{1},\hat{s}_{-}\right)
d\vec{r}_{1}
\end{split}
\end{align}

\noindent where $\kappa=\nu/c$ is wavenumber, and we have introduced the
receive intensity point spread function $\mathrm{PSF}_{r}\left(  \vec{r}\right)
=H_{r}\left(  \vec{r}\right)  H_{r}^{\ast}\left(  \vec{r}\right)  $. Note that this function  is inherently real independently of the presence of aberrations in the receive path. 

\textit{Effect of aberrator}

To proceed from Eq. \ref{DPC1}, we must evaluate $P\left(  \vec{r}_{1},\hat{s}%
_{+}\right)  P^{\ast}\left(  \vec{r}_{1},\hat{s}_{-}\right)  $. In the absence
of aberrations, this evaluation is straightforward and leads to a fringe
pattern localized about $\vec{r}_{1}$. On the other hand, in the presence of
aberrations, the plane waves become distorted. Importantly, they become
distorted differently depending on their direction, leading to an
aberration-dependent phase shift in the fringe pattern. We proceed now to
evaluate this distortion by assuming that the aberration in question is both
thin and weakly varying, and located at depth $z_{0}=0$ (corresponding to our
angular-compounding shear depth). The resulting plane wave throughout the
sample can be written as a boundary-value solution

\begin{equation}\label{P}
P\left(  \vec{r}_{1},\hat{s}\right)  =\int D\left(  \vec{r}_{1}-x_{0}\right)
e^{i2\pi\kappa c\tau_{a}\left(  x_{0}\right)  }P\left(  x_{0},\hat{s}\right)
dx_{0}
\end{equation}

\noindent where $P\left(  x_{0},\hat{s}\right)  $ is the plane-wave pulse
incident on the aberration (assumed aberration-free), $\tau_{a}\left(
x_{0}\right)  $ is the time delay caused by the aberration, and $D\left(
\vec{r}\right)  $ is the 2D homogeneous-space propagator. In other words,
$D\left(  \vec{r}\right)  $ propagates a 2D pressure wave from one depth to
another. We find then

\begin{align}
\begin{split}
&P\left(  \vec{r}_{1},\hat{s}_{+}\right)  P^{\ast}\left(  \vec{r}_{1},\hat
{s}_{-}\right)  =\iint D\left(  \vec{r}_{1}-x_{0}\right)  D^{\ast}\left(
\vec{r}_{1}-x_{0}^{\prime}\right)  \\
&\times e^{i2\pi\kappa c\left(  \tau_{a}\left(
	x_{0}\right)  -\tau_{a}\left(  x_{0}^{\prime}\right)  \right)  }P\left(
x_{0},\hat{s}_{+}\right)  P^{\ast}\left(  x_{0}^{\prime},\hat{s}_{-}\right)
dx_{0}dx_{0}^{\prime}
\end{split}
\end{align}

In what follows, we make several approximations. First, we assume that the
off-axis tilt angle $\theta$ of the incident plane-wave pulse is small,
allowing us to write

\begin{equation}
P\left(  x_{0},\hat{s}\right)  =P_{0}e^{-\left(  \kappa-\kappa_{0}\right)
	^{2}/2\kappa_{\sigma}^{2}}e^{i2\pi\theta\kappa x_{0}}
\end{equation}

\noindent where $\kappa_{0}=\nu_{0}/c$ and $\kappa_{\sigma}=\nu_{\sigma}/c$.
Second, we assume that off-axis tilt angles remain small, despite the presence
of the aberrator. This allows us to use the paraxial approximation for
$D\left(  \vec{r}\right)  $, given by 

\begin{equation}
D\left(  \vec{r}\right)  =\sqrt{\frac{\kappa}{iz}}e^{i2\pi\kappa z} 
e^{i\pi\kappa x^{2}/z}
\end{equation}

\noindent(this is modified from \cite{mast2007} and may be derived simply by integrating
the 3D paraxial propagator \cite{goodman2017} over all $y$).

At this point, it is convenient to make the coordinate transformation
$x_{d}=x_{0}-x_{0}^{\prime}$ and $x_{c}=\frac{1}{2}\left(  x_{0}+x_{0} 
^{\prime}\right)  $, leading to

\begin{align}
&dx_{0}dx_{0}^{\prime}\rightarrow dx_{c}dx_{d} \\
&D\left(  \vec{r}_{1}-x_{0}\right)  D^{\ast}\left(  \vec{r}_{1}-x_{0}^{\prime
}\right)  \rightarrow\frac{\kappa}{z_{1}}e^{-i2\pi\kappa x_{d}\left(
	x_{1}-x_{c}\right)  /z_{1}} \\
&P\left(  x_{0},\hat{s}_{+}\right)  P^{\ast}\left(  x_{0}^{\prime},\hat{s}
_{-}\right)  \rightarrow P_{0}^{2}e^{-\left(  \kappa-\kappa_{0}\right)
	^{2}/\kappa_{\sigma}^{2}}e^{i2\pi\kappa\left(  x_{d}\theta_{c}+x_{c}\theta
	_{d}\right)}  
\end{align}

We also make the approximation

\begin{equation}
\tau_{a}\left(  x_{c}+\tfrac{1}{2}x_{d}\right)  -\tau_{a}\left(  x_{c}%
-\tfrac{1}{2}x_{d}\right)  \approx x_{d}\partial_{x}\tau_{a}\left(
x_{c}\right) 
\end{equation}

\noindent where $\partial_{x}$ is shorthand for $d/dx$. We note that this
approximation requires only the slope of the aberration to be small, and not
the aberration itself, which is an advantage of our differential approach. We
also note that because of our angular compounding approach, we need only
consider a single pair of plane-wave pulses of tilt angles centered about zero
(i.e. $\theta_{c}=0$), since all other angles $\theta_{c}$ become essentially
``untilted" about the same center.

Combining these approximations, we obtain

\begin{align}\label{PP}
\begin{split}
P&\left(  \vec{r}_{1},\hat{s}_{+}\right)  P^{\ast}\left(  \vec{r}_{1},\hat
{s}_{-}\right)  =\frac{\kappa }{z_{1}}P_{0}^{2} e^{-\left(  \kappa-\kappa
	_{0}\right)  ^{2}/\kappa_{\sigma}^{2}}  \\
&\times \iint e^{-i2\pi\kappa x_{d}\left[
	\left(  x_{1}-x_{c}\right)  /z_{1}-c\partial_{x}t\left(  x_{c}\right)
	\right]  }e^{i2\pi\theta_{d}\kappa x_{c}}dx_{c}dx_{d}
\end{split}
\end{align}

\noindent which can be inserted into Eq. \ref{DPC1}, obtaining

\begin{align}\label{BB}
\begin{split}
&B_{+}\left(  \vec{r}\right)  B_{-}^{\ast}\left(  \vec{r}\right)  _{\nu}%
=\frac{\kappa }{z}\bar{\gamma}P_{0}^{2}e^{-\left(  \kappa-\kappa
	_{0}\right)  ^{2}/\kappa_{\sigma}^{2}} \\
&  \times \iint e^{-i2\pi\kappa x_{d}\left[
	\left(  x-x_{c}\right)  /z-c\partial_{x}t\left(  x_{c}\right)  \right]
}e^{-i2\pi\theta_{d}\kappa\left(  x-x_{c}\right)  }dx_{c}dx_{d}
\end{split}
\end{align}

\noindent where we have made the additional approximations that $\left(
\hat{s}_{+}-\hat{s}_{-}\right)  \cdot\vec{r}\approx\theta_{d}x$ and that the
aberrations are weak or localized enough that the receive PSF remains
reasonably localized.

Finally, Eq. \ref{BB} can be integrated over $\nu$ (or equivalently $\kappa$) to
yield a formal solution to $B_{+}\left(  \vec{r}\right)  B_{-}^{\ast}\left(
\vec{r}\right)  $, and hence ultimately to $\Delta\Phi\left(  \vec{r}\right)
$. However, the remaining double integral over finite ranges of $x_{c}$ and
$x_{d}$ is unwieldy and provides little insight. Instead, we follow a
different approach which, at the expense of accuracy, provides some degree of intuition.

In particular, let us consider the range of $x_{d}$'s that contribute to the
integral in Eq. \ref{BB}. At large depths, this range is limited by the size of the
transducer itself. At shallower depths, it is even more limited by the
wavenumber-dependent numerical aperture associated with the transducer element
size. In either case, we can introduce a phenomenological window for $x_{d}$
by introducing a bounding function $\exp(-\kappa^{2}\varphi_{\kappa}%
^{2}x_{d}^{2}$) into the integrand of Eq. \ref{BB}. The addition of this function
allows Eq. \ref{BB} to be readily integrated, first over $x_{d}$ and then over
$\nu$, leading to our final result given by Eq. \ref{PhiDPCforward}. 

We emphasize that our introduction of a bounding function for $x_{d}$ is
somewhat ad hoc. Nevertheless, it leads ultimately to a simple interpretation
of DPC, as discussed in Section \ref{forward}. In practice, we found that our numerical
simulations were reasonably robust and did not significantly depend on
$\varphi_{\kappa}$, which we chose to be typically around $0.05$. In general,
increasing $\varphi_{\kappa}$ led to an increased smoothing of our
simulations, particularly at large depths.

\bibliographystyle{IEEEtran}
\bibliography{DPCrefs}
\end{document}